\documentclass[twocolumn,showpacs,preprintnumbers,amsmath,amssymb,prb]{revtex4} \newcommand\PrePrint{0}
\usepackage[dvips]{color,graphics,epsfig,rotating}
\usepackage{graphicx}
\usepackage{dcolumn}
\usepackage{bm}
\usepackage{color}
 
\makeatletter
\ifthenelse{\equal{\PrePrint}{1}}
{
  \def\@dotsep{4.5}
}
{
}
\makeatother
 
\newcommand\eq[1]                              
{
\begin{align*}
#1
\end{align*}
}
 
\newcommand\eql[2] 
{
\begin{equation}\label{#1}
\begin{split}
#2
\end{split}
\end{equation}
}
 
\newcommand\eqsl[1]                            
{
\begin{align}
#1
\end{align}
}
 
\newcommand\Eq[1]      {Eq.~\eqref{#1}}
\newcommand\Eqs[1]     {Eqs.~\eqref{#1}}

\newcommand\Sec[1]     {Sec.~\ref{#1}}
\newcommand\Ref[1]     {Ref.~\onlinecite{#1}}
\newcommand\Refs[1]    {Refs.~\onlinecite{#1}}

\newcommand\ME[3]      {\langle{{#1}}|{{#2}}|{{#3}}\rangle} 
\newcommand\ket[1]     {|{{#1}}\rangle}

\newcommand\braket[2]  {\langle{{#1}}|{{#2}}\rangle}

\newcommand\PsiGS      {\Psi_0}
\newcommand\PsiT[1][]  {\Psi_{\mathrm{T}#1}^{}}

\newcommand\Half       {\frac{1}{2}}

\newcommand\kvec       {\mathbf{k}}

\newcommand\Bop        {{\hat{B}}}
\newcommand\Hop        {{\hat{H}}}
\newcommand\Kop        {{\hat{K}}}

\newcommand\Vop        {{\hat{V}}}
\newcommand\vop        {{\hat{v}}}
\newcommand\vvecop     {{\hat{\mathbf{v}}}} 
\newcommand\vMFvec     {\bar{\mathbf{v}}_{\textrm{mf}}} 

\newcommand\EL         {E_\mathrm{L}^{}}         
\newcommand\EH         {E_\mathrm{H}^{}}      

\newcommand\ABINIT     {{\footnotesize{ABINIT}}}
\newcommand\ELK        {{\footnotesize{ELK}}}

\newcommand\OPIUM      {{\footnotesize{OPIUM}}}

\newcommand\Ecut       {E_{\mathrm{cut}}}
\newcommand\Eh[1][]    {\ensuremath{E_\textrm{h}#1}}
\newcommand\mEh[1][]   {\textrm{m}\ensuremath{E_\textrm{h}#1}}

\newcommand\Order[1]   {\mathcal{O}\left(#1\right)}
 
\definecolor{Green}{rgb}{0.2,0.96,0.2}
\definecolor{Remarks}{rgb}{1,0.3,0.3}
\definecolor{Extra}{rgb}{0.2,0.2,1}
\definecolor{Blue}{rgb}{0.2,0.3,1}
\definecolor{Black}{rgb}{0,0,0}

\newcommand\COMMENTED[1] {}

\newcommand\FIGDIR[1]   {figs/}
 
\begin{document}
 
\title{
Pressure-induced diamond to $\beta$-tin transition in bulk silicon:\\      
a near-exact quantum Monte Carlo study
}
 
\author{Wirawan Purwanto}
\author{Henry Krakauer}
\author{Shiwei Zhang}
\affiliation{Department of Physics, College of William and Mary,
Williamsburg, Virginia 23187-8795, USA}

\date{\today}
 
\begin{abstract}
 
The pressure-induced structural phase transition from
diamond to $\beta$-tin in silicon is an excellent test for theoretical total
energy methods.  The transition pressure provides a sensitive measure
of small relative energy changes between the two phases (one a
semiconductor and the other a semimetal). Experimentally, the
transition pressure is well characterized. Density-functional results
have been unsatisfactory. Even the generally much more accurate
diffusion Monte Carlo method has shown a noticeable fixed-node
error. We use the recently developed phaseless auxiliary-field quantum
Monte Carlo (AFQMC) method to calculate the relative energy
differences in the two phases.
In this method, all but the error due to the phaseless constraint can
be controlled systematically and driven to zero.  In both structural phases we
were able to benchmark the error of the phaseless constraint by carrying out 
exact unconstrained AFQMC calculations for
small supercells. Comparison between the two
shows
that the systematic error in the absolute total energies due
to the phaseless constraint is well within
0.5m\Eh/atom. Consistent with these internal benchmarks, the 
transition pressure obtained by the phaseless AFQMC from large
supercells is in very good agreement with experiment.

\end{abstract}


\maketitle
 
\section{Introduction}
\label{sec:intro}
 
Theoretical and computational treatment of the effects of electron
correlations remains a significant challenge. Despite decades of
effort invested into solving the Schroedinger equation (by
independent-particle, mean-field and perturbative methods), there are
still major difficulties in predicting and explaining many phenomena
related to bonding, cohesion, optical properties, magnetic orderings,
superconductivity and other quantum effects.  The pressure-induced
structural phase transition in silicon from diamond to $\beta$-tin
\cite{Mujica2003} is an excellent test for theoretical total energy
methods.  The transition pressure provides a sensitive measure of
small relative energy changes between the two phases (one a
semiconductor and the other a semimetal). Experimentally, the
transition pressure is well characterized.  
Density-functional theory (DFT) results have been unsatisfactory, exhibiting 
sensitivity to the particular form of the exchange-correlation (xc) 
functional.  
Even the
generally much more accurate diffusion Monte Carlo (DMC) method
\cite{moskowitz:1982,Reyn82,Foulkes2001,QMC_book_1994,KW} has shown
\cite{Alfe2004} a noticeable fixed-node \cite{Ande76} error.

The phaseless auxiliary-field (AF) quantum Monte Carlo (QMC) AFQMC
method
\cite{SZ-HK:2003,Al-Saidi_GAFQMC_2006,QMC-PW-Cherry:2007}
provides a new alternative for
{\em ab initio} many-body calculations to address electron correlation
effects.
All stochastic QMC methods
\cite{Ceperley1980,Reynolds1982,Foulkes2001,SZ-HK:2003} use projection
from a reference many-body wave function.
In principle these methods are exact.  In practice, however, the
fermionic sign problem
\cite{Ceperley_sign,kalos91,Zhang1999_Nato,Foulkes2001,SZ-HK:2003}
causes exponential growth of the variance with system size and
projection time.  Transient methods,
\cite{Ceperley1980,Shiftcont1997,Shiftcont1998} which maintain
exactness while enduring the sign problem, can be very useful 
{\em if\/} sufficiently accurate information can be obtained with a
relatively short projection, as we illustrate in the present 
paper (Sec.~\ref{sec:benching-phaseless}).
In general, however, the sign problem 
must be completely eliminated (usually with an approximation)
to achieve
a general, efficient method for realistic systems.
The majority of QMC calculations in fermion systems 
have been done in this form, for example with the 
fixed-node approximation \cite{Ande76,Foulkes2001} in DMC, which has been 
the most commonly applied QMC method in electronic structure.
 
The phaseless AFQMC controls the sign problem
with a global phase condition in the over-complete manifold of Slater
determinants (in which antisymmetry is imposed). 
Since the antisymmetry ensures that each walker is automatically 
``fermionic'', the tendency for the walker population to collapse to a 
global bosonic 
state is eliminated in this approach. It is reasonable to expect that
an overall phase constraint applied in this manifold
to be less
restrictive\cite{Zhang1999_Nato}.
Applications indicate that this often is the case.
In a variety of systems AFQMC has demonstrated accuracy
equaling or surpassing the most accurate (non-exponential scaling)
many-body computational methods.
These include first- and second-row molecules, \cite{Al-Saidi_GAFQMC_2006,AFQMC-CPC2005} 
transition metal oxide molecules, \cite{Al-Saidi_TiO-MnO:2006} 
simple solids, \cite{SZ-HK:2003,2008-FS-Hendra} 
post-$d$ elements \cite{Al-saidi_Post-d:2006}
van der Waals systems, \cite{QMC-Hbonded_al-saidi:2007}
molecular excited states, \cite{2009-C2-exc-afqmc}
and in molecules in which bonds are being stretched or broken.
\cite{AFQMC-bondbreak-al-saidi:2007,2008-SpnContam-Purwanto,2009-C2-exc-afqmc}
Most of these calculations used a mean-field single determinant 
taken directly from DFT or Hartree-Fock (HF)
for the trial wave function in the phaseless constraint.
As a result, the phaseless AFQMC method 
reduces the reliance of QMC on the
quality of the trial wave function. 
\cite{Al-Saidi_GAFQMC_2006,AFQMC-bondbreak-al-saidi:2007,2008-SpnContam-Purwanto}
This is desirable in order to make QMC more of a general and ``blackbox''
approach.
 
The use of a basis set is a second feature that distinguishes the AFQMC method from the standard
DMC method.  \cite{moskowitz:1982,Reyn82,Foulkes2001,QMC_book_1994,KW}
The latter works in electron
coordinate space. As a result, there is no finite basis set error 
{\em per se\/} in DMC.  There are presently two main flavors of the phaseless
AFQMC method, corresponding to two different choices of the
one-electron basis: (i) planewave with norm-conserving pseudopotential
(as widely adopted in solid state physics),
\cite{SZ-HK:2003,QMC-PW-Cherry:2007} and (ii) Gaussian type basis sets
(the standard in quantum chemistry). \cite{Al-Saidi_GAFQMC_2006}
In  planewave AFQMC, convergence to the basis set limit is easily controlled, 
as in DFT calculations, using the plane wave cutoff energy $\Ecut$.
 
In this paper, planewave AFQMC is used to calculate the relative energy
differences between the two phases. 
The goal is to examine the accuracy of phaseless AFQMC,
benchmarking the energy difference at the transition volumes against 
experiment and
DMC results, and against exact free-projection 
AFQMC using smaller primitive cells.
In the phaseless AFQMC approach, all but the error
from the phaseless constraint can be controlled systematically and
driven essentially to zero.  
Comparison with exact AFQMC free-projection
shows that the systematic
error in the total energies due to the phaseless constraint is well within
0.5m\Eh/atom. Consistent with these internal benchmarks, the
transition pressure calculated from the phaseless AFQMC in large
supercells is found to be in very good agreement with experiment.
 
The paper is organized as follows. 
Several aspects of the AFQMC method,
including the hybrid formulation and the reduction of weight fluctuation, 
are described in Sec.~\ref{sec:method}. This is followed
by specific planewave AFQMC calculational details in Sec.~\ref{sec:details}.
Calculated results
are presented and discussed in Sec.~\ref{sec:rslt}. Finally, we
summarize and conclude in Sec.~\ref{sec:summary}.

\section{AFQMC methodology}
\label{sec:method}
 
This section reviews aspects of the AFQMC method in some detail.
This is done to facilitate the discussion of systematic errors in
Secs.~\ref{sec:details} and \ref{sec:rslt}, and to provide 
additional details on 
some phaseless AFQMC variants which are used in this
paper.  More complete descriptions of the phaseless AFQMC method can
be found in
\Refs{SZ-HK:2003,Purwanto2004,Purwanto2005,Al-Saidi_GAFQMC_2006,QMC-PW-Cherry:2007}. 
 
\subsection{AFQMC projection by random walks}
\label{sec:AFQMC}
 
The ground state of a many-body system, $\ket{\PsiGS}$, is obtained by
means of iterative projection from a trial wave function $\ket{\PsiT}$:
\eql{eq:gs-proj}
{
    e^{-\tau \Hop}
    e^{-\tau \Hop}
    \cdots
    e^{-\tau \Hop}
    \ket{\PsiT}
  \rightarrow \ket{\PsiGS}
    \,,
}
where $\Hop = \Kop + \Vop$ is the Hamiltonian of the system, consisting of
all one-body terms, $\Kop$, and two-body terms, $\Vop$.
AFQMC implements the ground-state projection as random walks in
the space of Slater determinants.
The Trotter-Suzuki breakup
\eql{eq:Trotter}
{
    e^{-\tau \Hop}
  = e^{-\tau \Kop / 2}
    e^{-\tau \Vop}
    e^{-\tau \Kop / 2}
  + \Order{\tau^3}
}
is used to separate the one- and two-body terms. Expressing $\Vop$ as a sum
of the squares of one-body operators $\{ \vop_i \} \equiv \vvecop$:
\eql{eq:Vop}
{
    \Vop
  = -\Half \sum_i \vop_i^2
  = -\Half \vvecop \cdot \vvecop
    \, ,
}
the Hubbard-Stratonovich (HS) transformation
\cite{Stratonovich1957,Hubbard1959}
is then used to express the two-body projector as a multidimensional integral
\eql{eq:HS-xform-noFB}
{
    e^{-\tau \Vop}
  = \prod_i \int_{-\infty}^{\infty} \frac{d\sigma_i}{\sqrt{2\pi}}
    e^{-\sigma_i^2 / 2}
    e^{\sqrt{\tau} \, \sigma_i \vop_i}
    \, .
}
Using Eq.~(\ref{eq:HS-xform-noFB}) effectively maps the two-body interaction 
onto
a fictitious non-interacting Hamiltonian with coupling to auxiliary classical 
fields  $\{{\sigma}_i\} \equiv \bm{\sigma}$.
The operation of the one-body projector on a Slater determinant $\ket{\phi}$ 
simply yields
another determinant: 
$\ket{\phi'}=e^{\sqrt{\tau} \, \bm{\sigma}\cdot \vvecop }\ket{\phi}$. 
If $\ket{\PsiT}$ in \Eq{eq:gs-proj} is expressed 
as a sum of Slater determinants ({\em e.g.}, just one if $\ket{\PsiT}$ 
is a HF or DFT solution), 
the integral in Eq.~(\ref{eq:HS-xform-noFB}) can then be evaluated 
using Monte Carlo sampling
over random walker streams. \cite{Zhang1997_CPMC,SZ-HK:2003}
 
As discussed further in Sec.~\ref{ssec:plAFQMC},
it is advantageous computationally to rewrite the two-body potential in 
Eq.~(\ref{eq:Vop}), subtracting
the mean-field contribution 
\cite{Shiftcont1997,Shiftcont1998,Purwanto2005,AlSaidi2006b}
prior to the HS transformation:
\eql{eq:V2b-mfbg}
{
    \Vop
& = -\left[
        \Half (\vvecop - \vMFvec)^2
        + \vvecop \cdot \vMFvec
        - \Half \vMFvec^2
    \right]
    \,,
}
where $\vMFvec$ is generally chosen to be the expectation value of the
$\vvecop$ operator with respect to the trial wave function 
\eql{eq:vbar-MF}
{
    \vMFvec
    \equiv
    \frac{\ME{\PsiT}{\vvecop}{\PsiT}}
         {\braket{\PsiT}{\PsiT}}
    \,.
}
 
\subsection{Phaseless AFQMC}
\label{ssec:plAFQMC}
 
In principle, the procedure 
in \mbox{Eqs.~(\ref{eq:gs-proj}-\ref{eq:HS-xform-noFB})} 
yields the exact ground state. The basic idea can be efficiently 
realized 
by branching random walks, as is used in Sec.~\ref{sec:benching-phaseless}
to carry out exact free-projection.
In practice, however, a phase problem appears, because the
repulsive Coulomb interaction gives
rise to imaginary $\vvecop$, complex walkers $\ket{\phi}$,
and complex $\braket{\PsiT}{\phi}$ overlaps,
causing the variance to grow exponentially and swamp the signal.
To control this problem, importance sampling and a phaseless 
approximation\cite{SZ-HK:2003} were
introduced, yielding a stable stochastic simulation.
The importance sampling transformation leads to a representation 
of the ground-state wave function as a weighted sum of Slater
determinants $\{ \ket{\phi} \}$: \cite{SZ-HK:2003,Purwanto2004}
\eql{eq:gs-wfn}
{
    \ket{\PsiGS}
  = \sum_{\phi} w_\phi \frac{\ket{\phi}}{\braket{\PsiT}{\phi}}
    \,.
}
A force bias term results in Eq.~(\ref{eq:HS-xform-noFB}):
\eql{eq:HS-xform}
{
    e^{-\tau \Vop}
  = \prod_i \int_{\infty}^{\infty} \frac{d\sigma_i}{\sqrt{2\pi}}
    e^{-\sigma_i^2 / 2}
    e^{\sigma_i \bar{\sigma}_i - \bar{\sigma}_i^2 / 2}
    e^{\sqrt{\tau} \, (\sigma_i - \bar{\sigma}_i) \vop_i}
    \,.
}
The corresponding importance-sampled one-body propagator then takes the form
\eql{eq:gs-iz-HS-proj}
{
    w_{\phi'} \ket{\phi'}
    \leftarrow
    \left[
        \int d\bm{\sigma} g(\bm{\sigma})
        \Bop(\bm{\sigma} - \bar{\bm{\sigma}})
        W(\bm{\sigma}, \bar{\bm{\sigma}})
    \right]
    w_{\phi} \ket{\phi}
    \,,
}
where $\{\bar{\sigma}_i\} \equiv \bar{\bm{\sigma}}$, 
$g(\bm{\sigma})$ is the multidimensional Gaussian probability density
function with zero mean and unit width, and
\eqsl
{
    \label{eq:B-def}
    \Bop(\bm{\sigma} - \bar{\bm{\sigma}})
&   \equiv
    e^{-\tau \Kop/2}
    e^{\sqrt{\tau} \, (\bm{\sigma} - \bar{\bm{\sigma}}) \cdot \vvecop}
    e^{-\tau \Kop/2}
    \,,
\\
    \label{eq:prop-imp}
    \ket{\phi'}
&   =
    \Bop(\bm{\sigma} - \bar{\bm{\sigma}})\,\ket{\phi}
    \,,
\\
    \label{eq:W-def}
    W(\bm{\sigma}, \bar{\bm{\sigma}})
&   \equiv
    \frac{\ME{\PsiT}{\Bop(\bm{\sigma} - \bar{\bm{\sigma}})}{\phi}}
         {\braket{\PsiT}{\phi}}
    e^{\bm{\sigma} \cdot \bar{\bm{\sigma}} -
       \bar{\bm{\sigma}} \cdot \bar{\bm{\sigma}} / 2}
    \,.
}
The one-body operator $\Bop$ generates the random walker stream, transforming
$\ket{\phi}$ into $\ket{\phi'}$, while $W$ updates the weight factor
$w_\phi$.
 
The optimal choice of $\bar{\bm{\sigma}}$, which cancels the weight
fluctuation to $\Order{\sqrt{\tau}}$, is given by
\eql{eq:sigmabar-optimal}
{
    \bar{\bm{\sigma}}
  = - \sqrt{\tau} \,
    \frac{\ME{\PsiT}{\vvecop}{\phi}}
         {\braket{\PsiT}{\phi}}
    \,,
}
where $\phi$ is the determinant being propagated.
Using this choice, the weight update factor $W$ 
can be written as \cite{SZ-HK:2003}
\eql{eq:W-EL}
{
    W(\bm{\sigma}, \bar{\bm{\sigma}})
    \approx
    e^{
        -\tau {\ME{\PsiT}{\Hop}{\phi}} / {\braket{\PsiT}{\phi}}
    }
    \equiv
    e^{
        -\tau \EL[\phi]
    }
    \,,
}
where $\EL[\phi]$ is referred to as the  ``local energy'' of $|\phi\rangle$.
In practice, we use the average of two local energies to update the
weight:
\eql{eq:W-EL-avg}
{
    W(\bm{\sigma}, \bar{\bm{\sigma}})
    \approx
    e^{
        -\tau (\EL[\phi'] + \EL[\phi]) / 2
    }
    \,.
}
The total energy can be calculated 
using the mixed-estimate form, which is not variational \cite{SZ-HK:2003}.
 
The key to controlling the phase problem is to prevent a two-dimensional
random walk in the complex $\braket{\PsiT}{\phi}$-plane,
thus avoiding the growth of a finite density at
the origin. To do this, the
phase rotation of the walker $\ket{\phi}$ is defined by
\eql{eq:phaserot}
{
    \Delta\theta
    \equiv
    \arg \left( \frac{\braket{\PsiT}{\phi'}}{\braket{\PsiT}{\phi}} \right)
    \,,
}
and the walker weight is ``projected'' to its real, positive value:
\eql{eq:phaseless-constraint}
{
    w_{\phi'}
    \leftarrow
    \begin{cases}
      \cos(\Delta\theta) \, |W(\bm{\sigma}, \bar{\bm{\sigma}})| \, w_{\phi}
      \,,
      & |\Delta\theta| < \pi/2
      \\
      0
      \,,
      & \textrm{otherwise}
    \end{cases}
    \,.
}
If the mean-field background is non-zero, its 
subtraction in Eq.~(\ref{eq:V2b-mfbg}) can lead to 
a reduction in the average rotation angle $\Delta\theta$ 
(and variance of the energy).
\cite{Purwanto2005,AlSaidi2006b}
 
\subsection{AFQMC in hybrid form}
 
Most applications to date have used the phaseless AFQMC local energy 
formalism, described above.
In planewave AFQMC, 
evaluating $\EL$ scales as $\Order{N^2M \log M}$, while 
the propagation step [\Eq{eq:gs-iz-HS-proj}] 
scales as $\Order{N M \log M}$, using
fast Fourier transforms.\cite{Suewattana2007}
Computation of the overlap matrix and other operations scale no 
worse than $\Order{N^2 M}$.
 
To reduce the frequency of evaluating $\EL$, the most costly part of the 
calculation, we can use an alternative formulation, 
the ``hybrid'' form \cite{SZ-HK:2003,Purwanto2004} of the walker weight in
Eq.~\eqref{eq:W-def}.
In the hybrid variant, only measurement evaluations of $\EL$ are needed.
Since the autocorrelation time is typically $50$-$100$ times the time step 
$\tau$, this
variant may be more efficient.
The hybrid method tends to have larger variance than the local energy 
method, however. 
The latter satistifies zero-variance in the limit of an exact $|\Psi_T\rangle$,
explicitly canceling out some $\Order{\tau}$ terms. 
The two methods also have different Trotter behaviors, 
as illustrated in Sec.~\ref{sec:Trotter-err}, but they approach the same 
answer as $\tau \to 0$.
The hybrid method is used for the large supercell calculations reported in this
paper.
 
\subsection{Random walk bounds: controlling rare event fluctuations}
\label{ssec:bounds}
 
For any finite population of walkers, the stochastic nature 
of the simulation does not preclude rare events, which cause 
extremely large population fluctuations.
For example, a walker near the origin of the
\mbox{$\braket{\PsiT}{\phi}$-plane} can acquire a very large weight 
in a move $|\phi'\rangle \leftarrow \Bop |\phi\rangle$ 
[\Eq{eq:gs-iz-HS-proj}], due to the occurence
of a very large 
$\braket{\PsiT}{\phi'} / \braket{\PsiT}{\phi}$ ratio [\Eq{eq:W-def}
or \Eq{eq:W-EL}].
To circumvent the problem in a simulation of finite population, 
we apply a bound condition in the local energy method:
\eql{eq:El-cap}
{
    (E_L^0 - \Delta\EL)  \le  \EL[\phi]  \le  (E_L^0 + \Delta\EL),
}
where the width of the energy range $\Delta E_L$ is defined as
\eql{eq:El-width}
{
    \Delta\EL \equiv \sqrt{\frac{2}{\tau}}
    \,,
}
and where the average local energy value $E_L^0$ 
is obtained by averaging $\EL$ measurements
during the growth phase.\cite{Zhang1997_CPMC}
If $\EL$ goes outside this range, it is capped 
at the maximum or minimum of the range.
For a typical $\tau$ ($\sim 0.05 \Eh^{-1}$), the energy range allowed by
\Eq{eq:El-cap} is large ($\sim 12 \Eh$), so $\EL$ is capped only in 
 very rare instances.

Similar bounds are introduced in the 
hybrid AFQMC method. Defining the hybrid energy as 
[compare Eqs.~(\ref{eq:W-def}] and 
(\ref{eq:W-EL}]
\eql{eq:Eh-def}
{
    \EH[\phi]
&   \equiv
    -\frac{\log W(\bm{\sigma}, \bar{\bm{\sigma}})}{\tau}
\\
& = -\frac{1}{\tau} \left[
    \log \left(
    \frac{\ME{\PsiT}{\Bop(\bm{\sigma} - \bar{\bm{\sigma}})}{\phi}}
         {\braket{\PsiT}{\phi}}
    \right)
    + \bm{\sigma} \cdot \bar{\bm{\sigma}}
    - \Half \bar{\bm{\sigma}} \cdot \bar{\bm{\sigma}}
    \right]
    \, ,
}
the value of $\EH$ is bounded as
\eql{eq:EH-cap}
{
    (E_H^0 - \Delta\EH)  \le  \EH[\phi]  \le  (E_H^0 + \Delta\EH)
    \, ,
}
where $E_H^0$ is estimated as in \Eq{eq:El-cap}.

In addition, the walker weights are also bounded such that 
$w_\phi \le w_{\max}$ at all
times for a reasonable $w_{\max}$ (typically set to the smaller of 
$100$ or $0.1$ times
the size of the population). This bound
is rarely triggered when the $\EH$ or $\EL$ bounding scheme is in place.
 
Finally, a force-bias bound is applied in both the local energy and hybrid 
methods.
This prevents
large modification of the orbitals when the denominator 
$\braket{\PsiT}{\phi}$ in
\Eq{eq:sigmabar-optimal} is small:
\eql{eq:FB-cap}
{
    |\bar{\sigma}_i| \le 1.0.
}
This bound is implicitly $\tau$-dependent, as seen in \Eq{eq:sigmabar-optimal}.
We have found that the energy cap ($\EL$ or $\EH$)
had the most effect in controlling
weight fluctuations.
 
It is important to note that 
the bounds being applied, while \emph{ad hoc\/}, have well-defined
limiting behavior. As $\tau \to 0$, the bounds on the physical quantities
$E_L$ and $\langle \vop\rangle$ both approach $\infty$.
The bounds only affect the Trotter error at finite $\tau$, but not the final 
answer when $\tau$ is extrapolated to zero. 
 
\subsection{Exact calculations: unconstrained AFQMC}

To estimate the accuracy of phaseless AFQMC, calculations using exact
unconstrained ``free'' projection were carried out (\Sec{sec:details}).
In free projection, the weights
$\{w_{\phi}\}$ are allowed to acquire a phase.
This is implemented using a modified form of the hybrid method,
where the mean-field average of the $\vvecop$ operators is used as the force 
bias
[instead of \Eq{eq:sigmabar-optimal}],
\eql{eq:sigmabar-MF}
{
    \bar{\bm{\sigma}}_{\textrm{mf}}
& = - \sqrt{\tau} \, \vMFvec \, .
}
This choice is equivalent to the subtraction of mean-field contribution to
the two-body potential described in \Eq{eq:V2b-mfbg}.
The use of the mean-field background subtraction is essential in
prolonging the stability of the simulation before 
the signal is lost to the phase problem.
None of the bounds in the preceding subsection is applied in the 
free-projection calculations.
 
\section{
AFQMC computational details for silicon diamond and $\beta$-tin
}
\label{sec:details}
 
The present calculations are carried out with planewave based AFQMC
(PW-AFQMC), which uses norm-conserving and separable Kleinman-Bylander
\cite{Kleinman1982} pseudopotentials to achieve efficient $\Order{N^2
M \log M}$ system size scaling, similar to planewave based DFT
(PW-DFT) calculations. We first describe specific computational details of the 
planewave AFQMC calculations, including the pseudopotential, 
planewave cutoff, and supercells.
 
Convergence to the basis set limit is easily
controlled, as in DFT calculations, using the plane wave cutoff energy
$\Ecut$.  Our calculations used $\Ecut = 12.5 \Eh$, which is the design 
cutoff of our Si 
pseudopotential (see below).
For material systems such as silicon, we have
previously shown \cite{QMC-PW-Cherry:2007} 
that a good $\Ecut$ at the DFT level, as determined by the norm-conserving 
pseudopotential, is sufficient to converge the two-particle 
correlations in AFQMC to within typical ststistical errors. 
In DFT calculations with the local density approximation (LDA),
the total absolute energies of the diamond primitive cell using this 
$\Ecut$ has an error of $\simeq 0.43\,\mEh$ (as verified by using increasingly
larger values of $\Ecut$). 
Basis set convergence errors of energy differences are much smaller, of course. 

AFQMC calculations for large 54-atom $3 \times 3 \times 3$ diamond and
$\beta$-tin supercells were done to obtain the transition pressures,
after finite-size corrections, discussed below.  
Test calculations,
such as pseudopotential tests and comparisons with benchmark exact
AFQMC, were carried out for the smaller 2-atom primitive unit cells.

For each supercell and $\kvec$-point, corresponding trial wave functions 
$\ket{\PsiT}$ were taken as generated from DFT-LDA, using the {\ABINIT} code \cite{Abinit}. 
In $\beta$-tin, 
random $\kvec$ points are used, rather than special points such as Monkhorst Pack sets,
to remove open-shell effects. 
For each $\kvec$, our single-determinant 
trial wave function 
is thus unique and non-degenerate at 
the ``Fermi surface.'' 
 
In the following subsections, aspects of the Si {\OPIUM} pseudopotential are first
discussed.  The quality of the pseudopotential is assessed by
comparing the equation of state (EOS) for the diamond and $\beta$-tin
structures with all-electron results within the framework of 
DFT. Next, efficient finite-size corrections are described, separately
analyzing one-body errors, which are analogous to $\kvec$ sampling in PW-DFT, and
two-body Coulomb finite size errors.

\begin{figure}[!htb]
\includegraphics[scale=0.33]{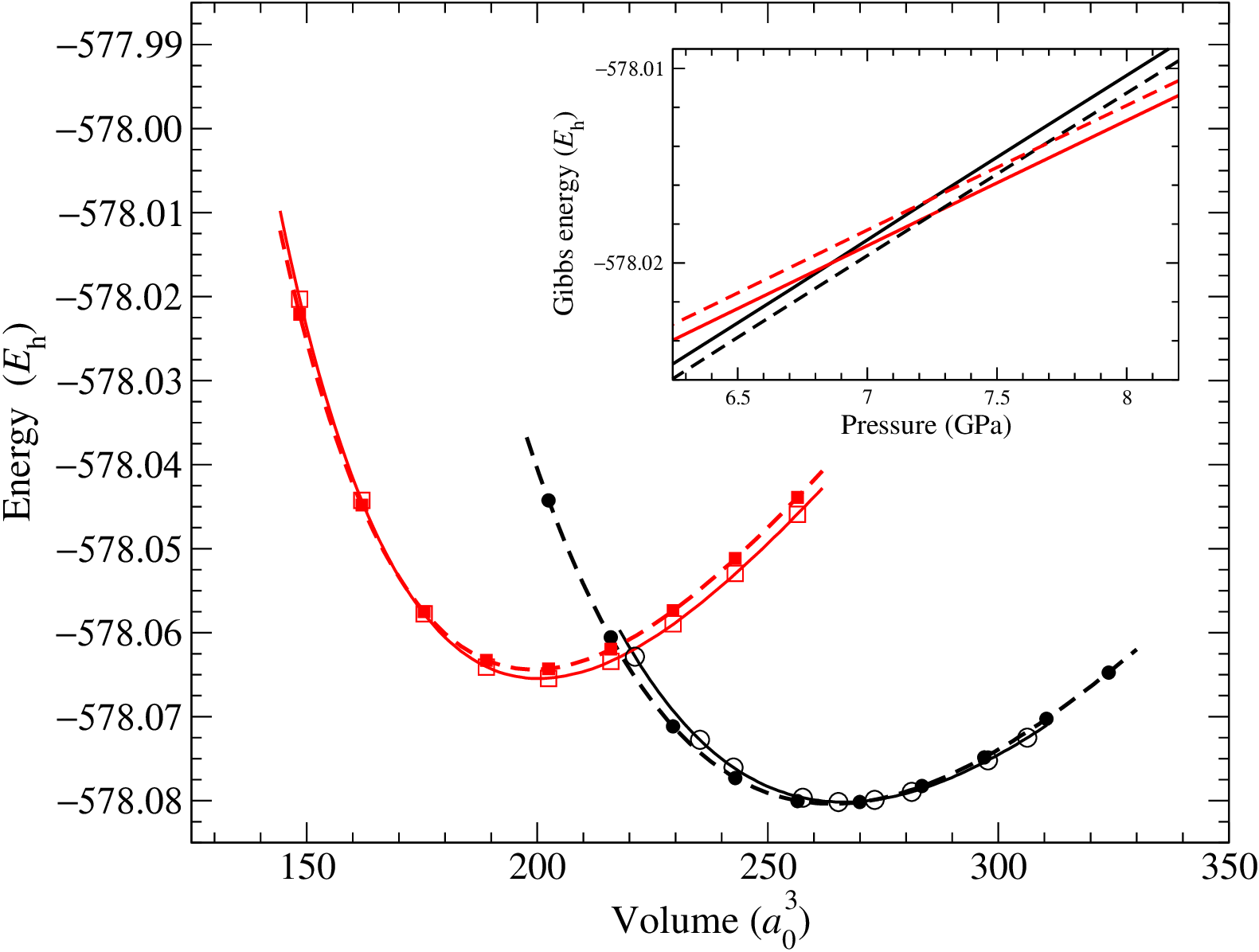}
\caption{
(Color online) Equation of state for the Si diamond and $\beta$-tin
 phases, comparing all-electron (solid lines) and pseudopotential (dashed lines) 
DFT-LDA results.
The inset shows the Gibbs energy as a function of the pressure.
}
\label{fig:DFT-EOS}
\end{figure}
 
\begin{table}[!htbp]
\caption{
Quantities of the diamond and $\beta$-tin phases of silicon
computed with DFT, using all-electron LAPW 
and planewave pseudopotential methods.
The xc functional used is the Perdew-Wang LDA. 
Volumes and lattice constants are expressed in atomic units ($a_0^3$ and
$a_0$, respectively);
energies are in eV;
bulk moduli and pressures are in GPa.
}
\label{tbl:DFT-obs}
 
\begin{ruledtabular}
\begin{tabular}{lcc}
Quantity                  & Pseudopotential & LAPW        \\
\hline
\textbf{diamond phase} \\
Equilibrium volume              & 263.888   & 266.474    \\
Equilibrium lattice constant    &  10.182   & 10.215     \\
Bulk modulus                    &  95.380   & 95.327     \\
Cohesive energy                 &   5.413   &  5.409     \\
\hline
\textbf{$\beta$-tin phase} \\
Equilibrium volume              & 199.132   & 200.364     \\
Bulk modulus                    & 114.760   & 114.947     \\ 
\hline
Transition pressure             &   7.67    &  6.86      \\
\end{tabular}
\end{ruledtabular}
\end{table}
 
\subsection{Si pseudopotential quality}
 
The optimized design method \cite{Rappe1990} was used to generate the Si pseudopotenital
with {\OPIUM}. \cite{OPIUM} The atomic reference state was
[Ne]\,3s$^2$\,3p$^{1.5}$\,3d$^{0.4}$.  All angular momentum channels
($l = 0, 1, 2$) used a cutoff radius $r_c = 2.08$~Bohr, with $l = 2$
as the local potential.  The optimized design 
pseudo-wavefunction was expanded using five spherical Bessel functions
with wave vector $q_c = 5.0 ~ \mathrm{Bohr}^{-1}$, which corresponds to a design
$\Ecut = 12.5 \, \Eh$, with a predicted planewave convergence error of
$1 \, \mEh$/atom for the absolute total energy. 
Explicit tests with DFT-LDA indicated errors several times 
smaller (see above), in both phases.
 
To test the quality of the pseudopotential, the EOS for
the diamond and $\beta$-tin structures was compared to all-electron
results within the framework of DFT. The results
are shown in Fig.~\ref{fig:DFT-EOS}.  All-electron 
calculations were done using the {\ELK} \cite{Elk} full-potential LAPW
program, and pseudopotential calculations with the planewave based {\ABINIT}\cite{Abinit} code
(using the same {\OPIUM} pseudopotential as in
AFQMC). The DFT-LDA Perdew-Wang \cite{Perdew1992} functional was
used.
In the all-electron and pseudopotential calculations,
identical dense $k$-point grids were used
($6 \times 6 \times 6$ in diamond and $16 \times
16 \times 16$ in $\beta$-tin).
A temperature broadening of $k_B T = 0.05$~eV was used in
the $\beta$-tin structure.  Birch-Murnaghan\cite{Birch1947} fits were
used to plot the Gibbs free energy.
The agreement for the EOS 
between the pseudopotential and all-electron calculations
is good, including the transition pressure values,
which differ by $\sim 0.8$\,GPa.
These results are quantified in Table~\ref{tbl:DFT-obs}.

\subsection{Trotter errors}
\label{sec:Trotter-err}
 
The transition pressure calculations in Sec.~\ref{sec:rslt} were done
for $3\times3\times3$ supercells, using a Trotter time step of $\tau =
0.025\,\Eh^{-1}$. In benchmarking exact AFQMC results in
Sec.~\ref{sec:benching-phaseless}, extraplotation to $\tau \to 0$ was
examined carefully for $1 \times 1 \times 1$ primitive cells for the phaseless
local-energy and hybrid AFQMC methods as well as for exact free-projection.
Not surprisingly, extrapolation
errors largely cancel between the two structures. For example, the
residual errors at $\tau = 0.025 \Eh^{-1}$ are $1.7(1)$ and $1.5(1)\,
 \mEh$ for diamond and $\beta$-tin primitive cells,
respectively. 
 
We also did several tests at larger supercell sizes. 
The residual error at $\tau = 0.025 \Eh^{-1}$
of a $3 \times 3 \times 3$ diamond structure supercell was estimated to be $1.6(4) \,
\mEh$ (normalized to the primitive cell), very similar to the value of $1.7(1)\,\mEh$ for the corresponding $1
\times 1 \times 1$ primitive cell. No explicit Trotter corrections were 
applied, therefore, in calculating the transition pressure, given the 
error cancellation between the two structures and the fact that the
estimated residual errors in even the {\em absolute\/} 
energy are not significantly larger than 
the QMC statistical errors.
 
\subsection{Finite-size errors}
\label{sec:FSerrors}
 
Independent-particle methods, such as DFT or HF, can
use Bloch's theorem to perform calculations in crystals, using only
the primitive unit cell. The macroscopic limit is achieved by
${\mathbf k}$-point quadrature in the Brillouin zone (BZ).  Many-body
methods, by contrast, must be performed for individual 
supercells.  The resulting finite-size (FS) errors often can be more
significant than statistical and other systematic errors. Eliminating
or reducing the FS errors is
crucial, therefore, to achieve accurate results.
The brute force extrapolations approach, using increasingly larger supercells, is
expensive and converges slowly, largely because two-body interactions
are long-ranged, causing FS effects to persist to large system sizes.
Alternatively, FS correction schemes can be used. 
\cite{Kent99a,Chiesa06,Kwee2008} 
 
\begin{figure}[ht]
\includegraphics[scale=0.3]{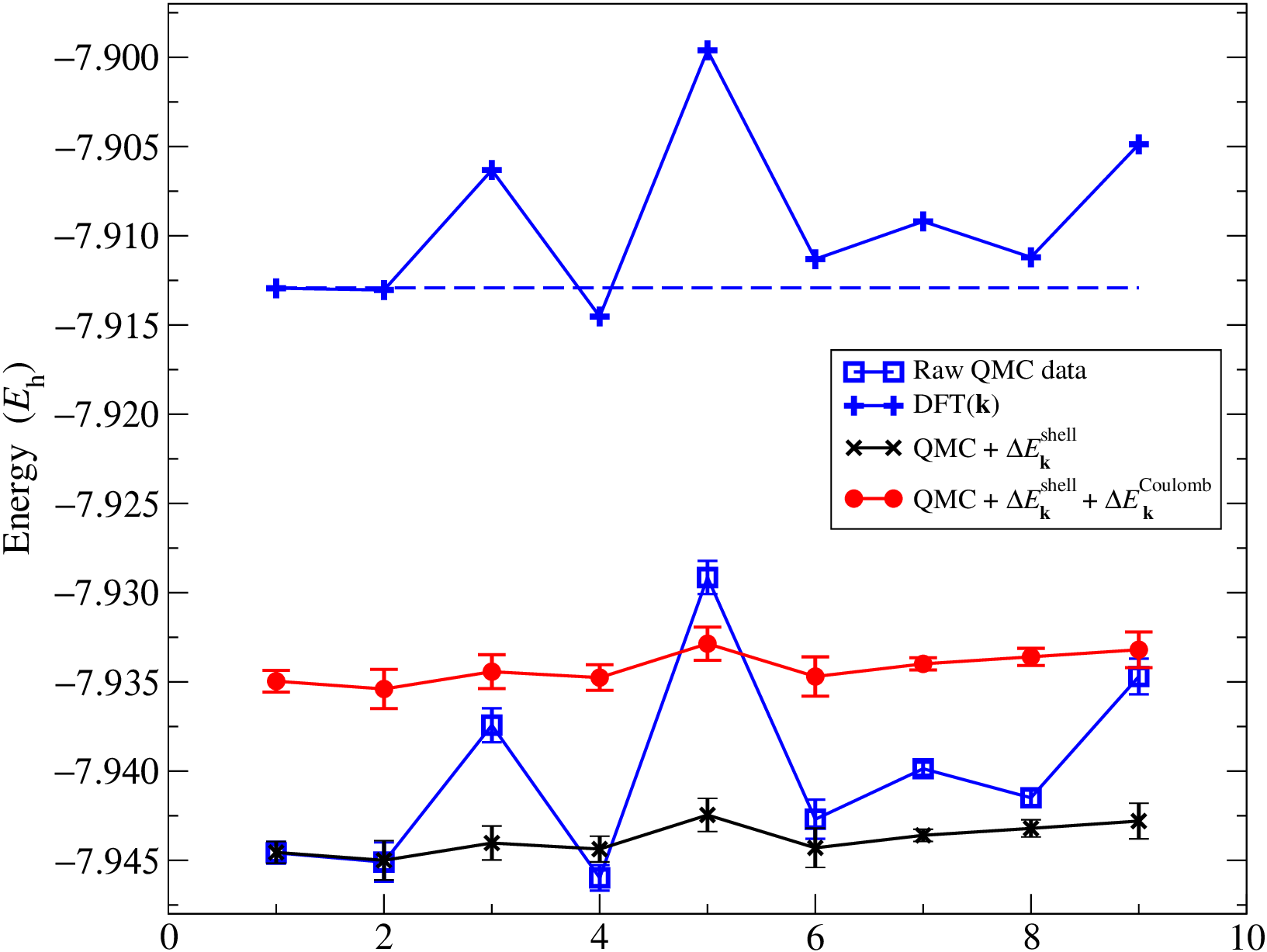}
\caption{
(Color online) AFQMC and LDA Si $\beta$-tin energies
at nine randomly-chosen $\kvec$ points in the Brillouin zone of
 $3 \times 3 \times 3$ supercell.
The AFQMC and LDA one-body FS errors are correlated, and the correction 
in \Eq{eq:FS-1B} reduces the variation in the QMC data. 
The two-body FS error is significant even with a 54-atom supercell, but 
is essentially independent of $\kvec$-points.
\label{fig:bSn-AFQMC-kpts}
}
\end{figure}

Both one- and two-body FS corrections \cite{Kent99a,Kwee2008} 
must be applied to achieve efficient convergence. 
One-body effects are related to BZ $\kvec$-point
sampling. These can be largely corrected, 
using DFT calculations to estimate
quadrature errors. In metals such at $\beta$-tin, BZ intergration errors are aggravated
by open-shell effects. Twist-averaged boundary conditions \cite{Lin2001,Kwee2008} 
can be used in this case to further reduce residual one-body errors, as is done here 
for the $\beta$-tin phase. 
The one-body FS correction is given by \cite{Kwee2008} 
\eql{eq:FS-1B}
{
    \Delta E^{\textrm{shell}}_\kvec
  = E^{\textrm{DFT}}
    - E^{\textrm{DFT}}_{\kvec} \, ,
}
namely by subtracting the DFT energy
at the same $\kvec$ vector ($E^{\textrm{DFT}}_{\kvec}$) and adding the DFT energy obtained with a
 dense $\kvec$ grid ($E^{\textrm{DFT}}$). 
Figure~\ref{fig:bSn-AFQMC-kpts} shows the reduced variation of the AFQMC total
energy after this correction is applied, for $3 \times 3 \times 3$ $\beta$-tin supercell. 
Averaging over the 9 randomly chosen $\kvec$ points before the correction 
results in a statistical error (combined error of the nine random
data points each of which has a statistical error bar) 
of $1 \, \mEh$, while averaging
after the correction reduces the combined error to $0.6 \, \mEh$.
As mentioned, 
random $\kvec$ points rather than special points were used 
to remove open-shell effects in metals and ensure that the trial wave function
is non-degenerate.  
 
The two-body FS error 
comes from the artificially induced periodicity
of the long-range electron-electron Coulomb repulsion, due to the use of periodic
boundary conditions.
This error can be reduced significantly, using the post-processing correction scheme of
Kwee {\em et al.},\cite{Kwee2008} 
which is based on a finite-size DFT xc functional,
corresponding to the finite-sized supercell. This two-body correction is given by
\eql{eq:FS-2B}
{
    \Delta E^{\textrm{Coulomb}}_\kvec
  = E^{\textrm{DFT}, \infty}_{\kvec}
    - E^{\textrm{DFT}, L}_{\kvec}
}
where $E^{\textrm{DFT}, \infty}_{\kvec}$ [$=E^{\textrm{DFT}}_{\kvec}$ 
in Eq.~(\ref{eq:FS-1B})] is the DFT energy computed with
the usual LDA xc functional, while $E^{\textrm{DFT}, L}_{\kvec}$
is the DFT energy computed with the {\em finite-size\/} LDA xc
functional.\cite{Kwee2008}
The $\kvec$-dependence of $\Delta E^{\textrm{Coulomb}}_\kvec$ is very small
compared to that of the one-body correction shown in Fig.~\ref{fig:bSn-AFQMC-kpts},
with variations of $\simeq 0.1 \mEh$ in $\beta$-tin.
 
The total FS correction is the result of 
applying the one- and two-body correction terms, \Eqs{eq:FS-1B} and \eqref{eq:FS-2B}, 
respectively. 
This is of course equivalent to applying 
$    \Delta E_\kvec
  = E^{\textrm{DFT}}
    - E^{\textrm{DFT}, L}_{\kvec}$
to the raw AFQMC energies. The corrected energies are averaged over the 
$\kvec$ points. 
The net effect of applying both FS corrections is to decrease the energy difference at 
the transition volumes from  34(1) to 29(1) $\mEh$. 
With these combined FS corrections, the residual errors in the 
absolute energies from $3 \times 3 \times 3$ supercells are expected to be
small in silicon. \cite{Kwee2008}
Error cancellation in
the energy difference between $\beta$-tin and diamond structures 
further reduces the error in the calculated transition pressure.

\section{Results and Discussion}
\label{sec:rslt}
 
\subsection{Benchmarking the phaseless approximation with exact free-projection AFQMC}
\label{sec:benching-phaseless}
 
The fermionic sign/phase constraints used by QMC methods generally
introduce uncontrolled approximations.
Examples include the DMC fixed-node approximation and AFQMC phaseless constraint.
Except where benchmarks with exact methods or experiment are available for comparison,
the corresponding constraint errors are difficult to quantify.
In this section, 
we show that exact free-projection
calculations are feasible for the primitive diamond and $\beta$-tin
structures, using planewave AFQMC on a large parallel computing platform.  
Comparison with the corresponding
approximate phaseless AFQMC calculations shows that the systematic
error due to the phaseless constraint is small (within 0.5 \mEh\,/atom), as 
described below.
 
\begin{figure}
\includegraphics[scale=0.33]{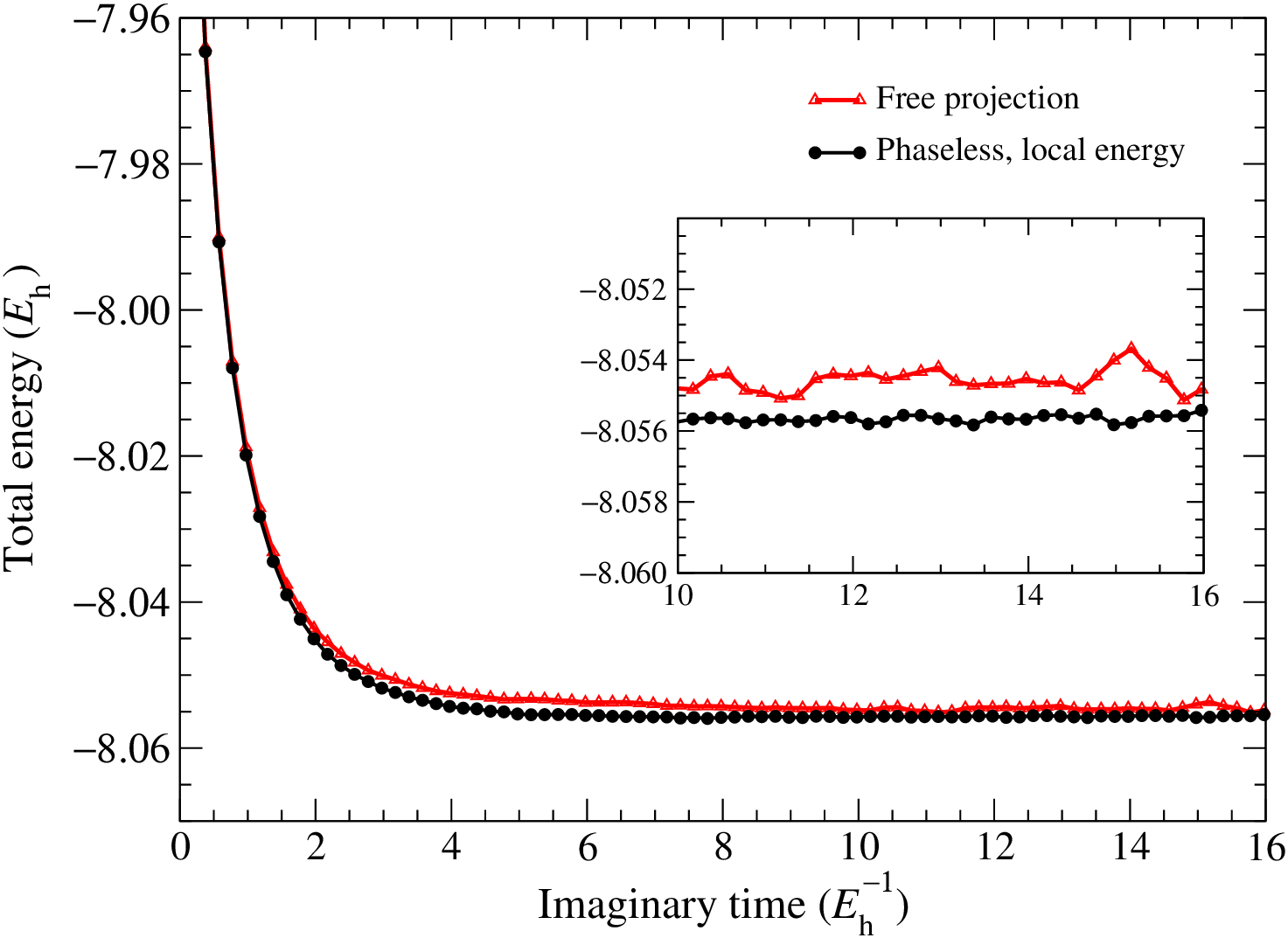}
\caption{
(Color online) Comparison of phaseless AFQMC with exact free-projection. 
The calculations shown are 
for the primitive cell in the diamond structure at the 
experimental equilibrium lattice constant ($10.264\,a_0$). 
The calculations used $L$ as the $\kvec$-point,
and a time step size of $\tau = 0.025 \Eh^{-1}$.
}
\label{fig:Si-111-eqlb}
\end{figure}

As illustrated in Fig.~\ref{fig:Si-111-eqlb} for the diamond structure,
free-projection to the ground state can be achieved in the primitive cell using large walker populations.
The free-projection calculation was done with a target population size of two million walkers, using about 2000 cores at the
NCCS Jaguar XT4 computer at Oak Ridge National Laboratory. 
An acceptable signal-to-noise ratio is sustained for sufficiently long imaginary
times. 
For projection times $\beta > \sim 16 \Eh^{-1}$, however, 
growing fluctuations, due to the phase problem, begin to emerge. 
Eventually the fluctuations become severe enough
to destroy the Monte Carlo signal.\cite{SZ-HK:2003}
The energy measurement for this benchmark is taken 
after the walkers are sufficiently equilibrated, $\beta > 10 \Eh^{-1}$. 
Similar calculations were performed in the $\beta$-tin structure.
 
\begin{figure}
\includegraphics[scale=0.33]{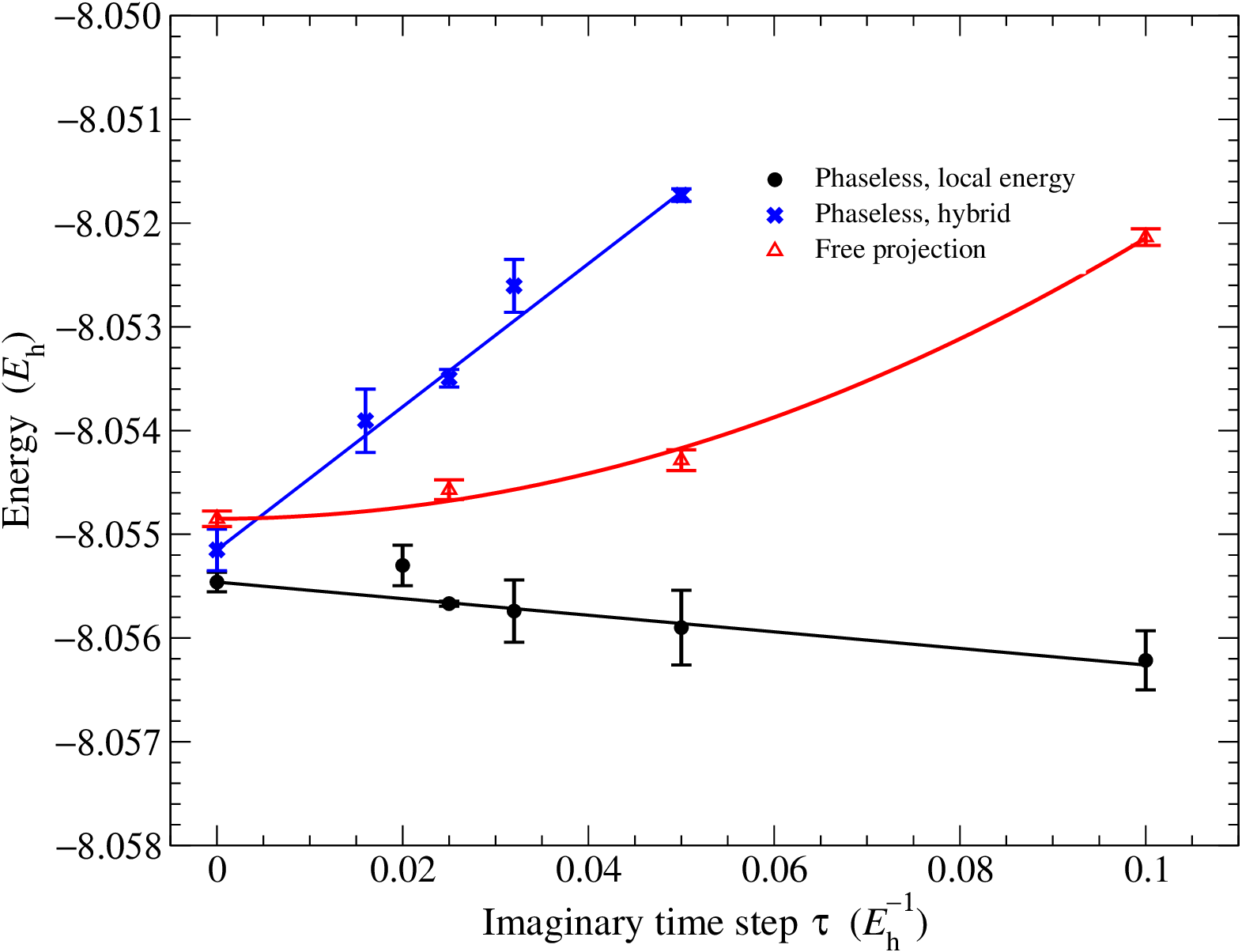}
\caption{
(Color online) Trotter time step $\tau$ extrapolation for the diamond structure,
as in Fig.~\ref{fig:Si-111-eqlb}, comparing local-energy and 
hybrid phaseless methods with free-projection.
}
\label{fig:Si-111-deltau}
\end{figure}

\begin{figure}
\includegraphics[scale=0.33]{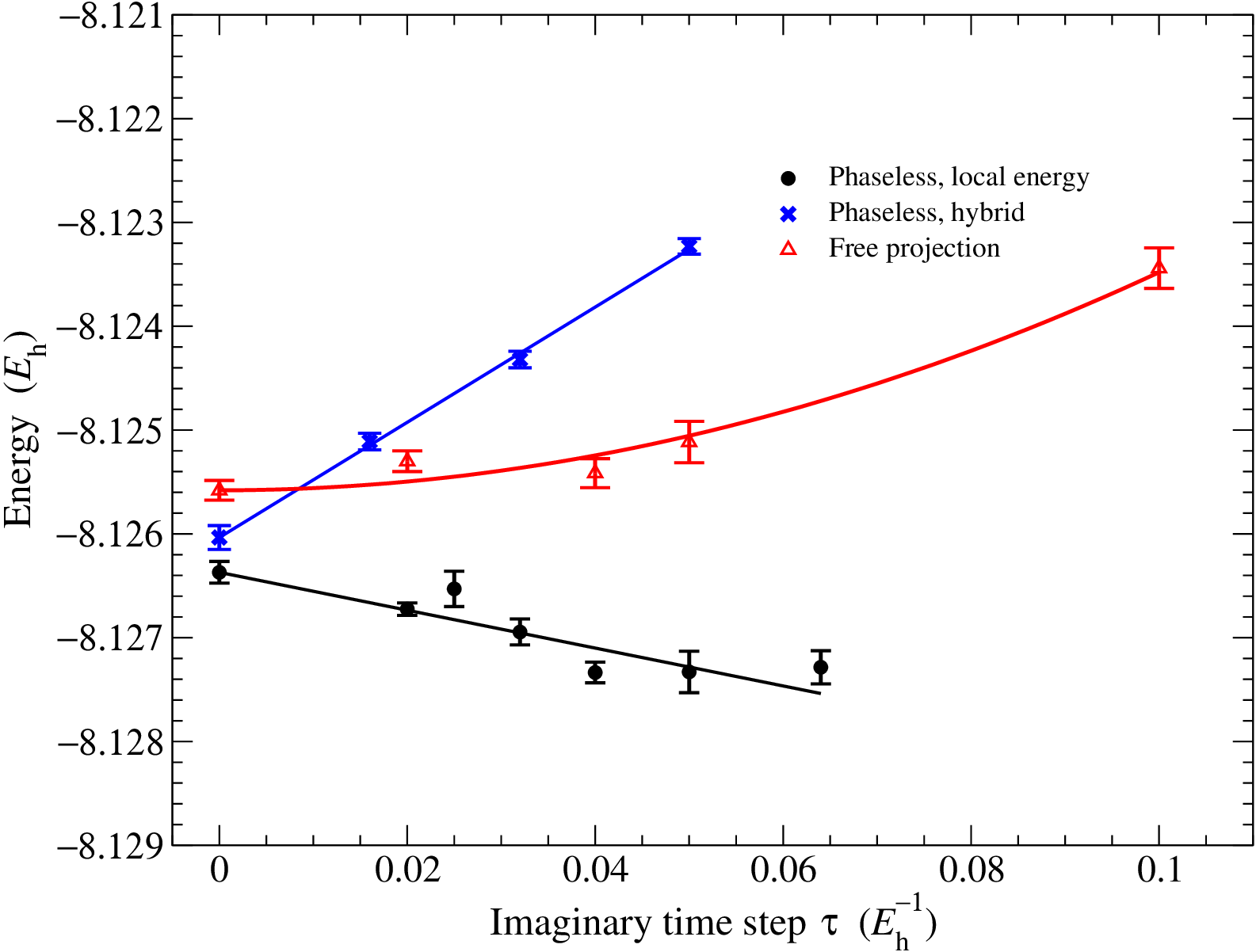}
\caption{
(Color online) Trotter time step $\tau$ extrapolation for the $\beta$-tin structure,
comparing the local-energy phaseless method with free-projection.
Calculations are for the primitive unit cell with volume of $30$ \AA$^3$
and $c/a=0.552$
at the reduced $\kvec$ = (0.25, 0.25, 0.25).
}
\label{fig:bSi-111-+25+25+25-eqlb}
\end{figure}

Figures~\ref{fig:Si-111-deltau} and \ref{fig:bSi-111-+25+25+25-eqlb} 
display extrapolations of the Trotter time step, $\tau$, for the diamond and
$\beta$-tin structures, respectively. 
The figures compare the local-energy and hybrid phaseless methods with
free-projection results.
As $\tau \to 0$, the
local-energy and hybrid phaseless methods 
are seen to converge to the same
result, 
as expected, but with different slopes.
To leading order, free-projection shows $\tau^2$ behavior,  while 
the local-energy and hybrid methods have linear $\tau$
behavior, since the phaseless constraint 
(and the bounds in Sec.~\ref{ssec:bounds}) in the latter two methods
break the quadratic scaling in Eq.~(\ref{eq:Trotter}).
The
hybrid method in Figs.~\ref{fig:Si-111-deltau} and \ref{fig:bSi-111-+25+25+25-eqlb} 
is seen to have the largest slope.
The Trotter behaviors of the respective methods are similar 
in the diamond and $\beta$-tin structures.
 
The error in the total energy caused by the phaseless
approximation, after extrapolation to $\tau \to 0$, is about $0.7\mEh$
(or $0.35 \mEh$ per atom) for diamond and $0.8\mEh$ in $\beta$-tin.
Note that the energy calculated from the phaseless approximation
using the mixed-estimate [Eq.~(\ref{eq:W-EL})] is not variational 
\cite{SZ-HK:2003}.
Indeed in both cases above it is 
below the exact result.
 
\begin{table}[!htp]
 
\caption{\label{tbl:ref-AFQMC}
AFQMC energies of Si diamond and $\beta$-tin phases for the 2-atom 
primitive cells, with volumes of $40.07$~\AA$^3$ and
$30.00$~\AA$^3$, respectively.
The $\beta$-tin $c/a$ ratio was set to 0.552.
Phaseless AFQMC values are from the local energy formalism.
Energies are in $\Eh$ units.
}
\begin{ruledtabular}
\begin{tabular}{llll}
phase       &reduced $\kvec$ vector& free projection & phaseless \\
\hline
diamond     & $(0.5, 0.5, 0.5)$    & $-8.05485(8)$   & $-8.0555(1)$ \\
$\beta$-tin & $(0.25, 0.25, 0.25)$ & $-8.1256(1)$    & $-8.1264(1)$ \\
$\beta$-tin & $(0.25,-0.25, 0.25)$ & $-8.0023(2)$    & $-8.0017(2)$ \\
\end{tabular}
\end{ruledtabular}
 
\end{table}
 
In Table~\ref{tbl:ref-AFQMC}, we list the absolute energies for three 
cases from our free-projection calculations, which can serve as benchmarks
in the future.
All the energies have been extrapolated to $\tau \to 0$.
The corresponding phaseless results are also listed, which are in
very good agreement with the exact results. We note that the 
two twist
boundary conditions ($\kvec$ points) in $\beta$-tin show different
behaviors. In $\kvec = (0.25, 0.25,
0.25)$ the phaseless energy is below the exact value (as in the diamond case),
while in the other, the phaseless energy is above. This manifests the 
varying quality of the trial wave function at the different $\kvec$-points, 
which shift the single-particle energy levels differently. 
Also, the FS effects are clearly very large in these small cells, where
the energy for the first $\kvec$-point is in fact below that of the diamond 
result.
 
\subsection{Transition pressure}
 
AFQMC calculations were done for $3 \times 3 \times 3$ supercells
(containing 54 silicon atoms), at the experimental transition volumes,
\cite{McMahon1994} $36.30$~\AA$^3$ for the diamond and $27.91$~\AA$^3$ 
for the $\beta$-tin structures.  For $\beta$-tin, we used 
the experimental value of $c/a =
0.550$. \cite{Hu1986}
(It was shown in \Ref{Alfe2004} that the dependence on $c/a$ is 
weak.)
Twist averaged boundary conditions used the single
special $\kvec$-point of Baldereschi \cite{Baldereschi1973} for the
diamond structure, and nine random $\kvec$ points for the
$\beta$-tin phase. 
The total energies lead to a ``raw'' transition pressure of $15.1(3)$\,GPa.
 
To compare with experiment, corrections are required to account for zero-point
motion and thermal effects. We apply these corrections as given in 
\Ref{Alfe2004}:
1) a zero-point motion
lowering of 0.3 GPa; 2) a room-temperature quasi-harmonic estimate of
the relative stabilization of the $\beta$-tin phase, which lowers the
pressure by 1.15 GPa.
This would give a transition pressure of $13.6(3)$\,GPa. 
In addition, 
standard mean-field pseudopotentials generated from
LDA or HF, such as the one used in the present paper, do not account for 
many-body effects in the core.
A correction was estimated in \Ref{Alfe2004}, by explicitly including a
many-body core polarization potential (CPP), which further lowers the 
pressure by $\sim 1.2$\,GPa. Assuming that our LDA pseudopotential is similar
to that in \Ref{Alfe2004}, we apply the same correction. 
Table~\ref{tbl:Si-Pt} reports our final result, and 
compares it to experiment and 
to other theoretical results. Corrections (1) and (2) have also been 
applied to the DFT results. 
 
\begin{table}[!hbtp]
 
\caption{\label{tbl:Si-Pt}
The transition pressure $P_t$ of the 
diamond to $\beta$-tin transition in silicon.
The AFQMC result is listed together with experimental and 
other theoretical results. To compare with experiment, appropriate
corrections have been applied to the theoretical results (see text).
}
\begin{ruledtabular}
\begin{tabular}{ll}
Method                         & $P_t$ (GPa) \\
\hline
LDA\cite{Needs1995}
                               & 6.7     \\
GGA (BP)\cite{Moll1995}
                               & 13.3  \\
GGA (PW91)\cite{Moll1995}
                               & 10.9  \\
DMC\cite{Alfe2004}
                               & 16.5(5)   \\
\textbf{AFQMC}   & \textbf{12.6(3)}  \\
Experiment\cite{Mujica2003}
                               & 10.3--12.5  \\
\end{tabular}
\end{ruledtabular}
 
\end{table}
 
The transition pressure is not very sensitive to the choice of
transition volumes.  For example, using the DMC predicted volumes
instead of the experimental values changes the energy difference by
only $0.01$ eV, from $0.49$ to $0.50$ eV, reducing the transition
pressure by less than $0.3$ GPa. 
 
The best calculation to date with the highest level of theory is 
the DMC calculations in \Ref{Alfe2004}. 
Compared to experiment, the somewhat overestimated DMC $P_t=16.5(5)$
value was attributed to the fixed-node error.\cite{Alfe2004}  This
seems consistent with our results. The DMC
discrepancy corresponds to a larger ``raw'' energy difference of $19.2
\, \mEh$ between the two phases, 
compared to $15.1(3) \, \mEh$ for phaseless AFQMC. As shown
in the previous subsection, the error due to the phaseless
approximation ($\simeq 0.5 \, \mEh$) appears to be an order of magnitude smaller
than this. 
Our calculations show that experiment and theory are in quantitative 
agreement on the 
diamond to $\beta$-tin transition.

\section{Summary}
\label{sec:summary}
 
We have applid the phaseless auxiliary-field quantum Monte Carlo method to 
study the pressure-induced structural phase transition from
diamond to $\beta$-tin in silicon.
This is a recently developed non-perturbative, 
many-body approach which recovers electron 
correlation by explicitly summing over fluctuating mean-field solutions
with Monte Carlo.
The only source of error which can not be systematically driven to zero
is that of the global phase constraint, used to control the sign/phase 
problem.
We quantified the systematic error from this phaseless approximation by
exact unconstrained AFQMC calculations in the primitive cell, carried out
on large parallel computers.
In both structural phases the error was found to be 
well within 0.5m\Eh/atom. A transition pressure was calculated form the
energy difference between the two phases at the experimental transition pressure, 
using 54-atom supercells. 
Twist-averaging boundary condition and finite-size corrections were
applied, which greatly accelerates the convergence to the thermodynamic 
limit. After corrections for zero-point effect, thermal effect, and 
the (lack of) core-polarization in the pseudopotential, 
the AFQMC results yield a transition pressure of $12.6\pm0.3$\,GPa, 
compared to experimental values of $10.3$-$12.5$\,GPa.
 
The good agreement between the phaseless AFQMC result and experiment is 
consistent with the internal benchmark with unconstrained AFQMC. 
Our analysis indicates
that the possible combined error from the 
calculations 
should be below $1$\,GPa. These include
pseudopotential transferability errors and core-polarization effect,
residual finite-size errors, 
and the error from the phaseless 
approximation.
 
\begin{acknowledgments}
 
The work was supported in part by DOE (DE-FG05-08OR23340 and
DE-FG02-07ER46366). H.K.~also acknowlesges support by ONR
(N000140510055 and N000140811235), and W.P.~and S.Z.~by NSF (DMR-0535592). 
Calculations were performed with support from INCITE at
the National Center for Computational Sciences
at Oak Ridge National Laboratory,
the Center for Piezoelectrics by Design,
and
the College of William \& Mary's SciClone cluster.
We are grateful to Eric Walter for many useful discussions and for
providing the pseudopotential used in this calculation.
 
\end{acknowledgments}

\end{document}